\documentclass{article}
\usepackage[dvips]{graphicx}  
\usepackage{color}

\usepackage[varg]{txfonts}
\usepackage{concmath}

\begin{document}

\newcommand{\rum}{\rule{0.5pt}{0pt}}
\newcommand{\rub}{\rule{1pt}{0pt}}
\newcommand{\rim}{\rule{0.3pt}{0pt}}
\newcommand{\numtimes}{\mbox{\raisebox{1.5pt}{${\scriptscriptstyle \rum\times}$}}}
\newcommand{\numtimess}{\mbox{\raisebox{1.0pt}{${\scriptscriptstyle \rum\times}$}}}
\newcommand{\Boldsq}{\vbox{\hrule height 0.7pt
\hbox{\vrule width 0.7pt \phantom{\footnotesize T}%
\vrule width 0.7pt}\hrule height 0.7pt}}
\newcommand{\two}{$\raise.5ex\hbox{$\scriptstyle 1$}\kern-.1em/
\kern-.15em\lower.25ex\hbox{$\scriptstyle 2$}$}

\renewcommand{\refname}{References}
\renewcommand{\tablename}{\small Table}
\renewcommand{\figurename}{\small Fig.}
\renewcommand{\contentsname}{Contents}

\twocolumn[%
\begin{center}
{\Large\bf 
One-Way Speed of Light Measurements Without Clock Synchronisation \rule{0pt}{13pt}}\par

\bigskip
Reginald T. Cahill \\ 
{\small\it  School of Chemical and Physical  Sciences, Flinders University,
Adelaide 5001, Australia\rule{0pt}{15pt}}\\
\raisebox{+1pt}{\footnotesize E-mail: Reg.Cahill@flinders.edu.au}\par

\bigskip

{\small\parbox{11cm}{%
The 1991 DeWitte double one-way 1st order in $v/c$  experiment   successfully measured the anisotropy of the speed of light   using clocks at each end of the RF coaxial cables.   However Spavieri {\it et al.}, Physics Letters A (2012), dpi: 10.1016/j.physleta 2012.0010, have reported that (i) clock effects caused by clock transport should be included, and (ii) that this additional effect cancels the one-way light speed  timing effect, implying that one-way light speed  experiments ``do not actually lead to the measurement of the one-way speed of light or determination of the absolute velocity of the preferred frame".  Here we explain that the Spavieri {\it et al.} derivation makes an assumption that is not always valid: that  the propagation is subject to the usual Fresnel drag effect, which is not the case for RF coaxial cables.  As well DeWitte did  take account of the clock transport effect. The Spavieri {\it et al.} paper has prompted this  clarification of  these issues.
\rule[0pt]{0pt}{0pt}}}\medskip
\end{center}]{%

\setcounter{section}{0}
\setcounter{equation}{0}
\setcounter{figure}{0}
\setcounter{table}{0}



\section{Introduction}  

The enormously significant 1991 DeWitte \cite{DeWitte} double one-way 1st order in $v/c$  experiment   successfully measured the anisotropy of the speed of light   using  clocks at each end of the RF coaxial cables. The technique uses rotation of the light path to permit extraction of   the light speed anisotropy, despite the clocks not being synchronised.  Data from  this 1st order in $v/c$  experiment  agrees  with the speed and direction of the anisotropy results from  2nd order in $v/c$ Michelson gas-mode interferometer experiments by Michelson and Morley and by Miller, see data in \cite{CahillGW2}, and with NASA spacecraft earth-flyby Doppler shift data \cite{CahillNASA}, and also with more recent 1st order in 
$v/c$ experiments  using a new single clock technique. 
  However Spavieri {\it et al.} \cite{Spavieri} reported that (i) clock effects caused by clock transport should be included, and (ii) that this additional effect cancels the one-way light speed  timing effect, implying that one-way light speed  experiments ``do not actually lead to the measurement of the one-way speed of light or determination of the absolute velocity of the preferred frame".  Here we explain that the Spavieri {\it et al.} derivation makes an assumption that is not always valid:  that  the propagation is subject to the usual Fresnel drag effect, which is not the case for RF coaxial cables.  The Spavieri {\it et al.}  paper has prompted this  clarification of  these issues.  In particular DeWitte took account of both the clock transport effect, and also that the RF coaxial cables did not exhibit a Fresnel drag, though these aspects were not discussed in \cite{DeWitte}.
  
\section{First Order in $\bf v/c$ Speed of EMR Experiments}
Fig.\ref{fig:oneway} shows the arrangement for measuring the one-way speed of light, either in vacuum, a dielectric, or RF coaxial cable.  It is usually argued that one-way speed of light measurements are not possible because the clocks $C_1$ and $C_2$ cannot be synchronised.   However  this is false, although an important previously neglected effect that needs to be included is the clock offset effect caused by transport  when the apparatus is  rotated \cite{Spavieri}, but most significantly the Fresnel drag effect is not present in RF coaxial cables.. 
 In   Fig.\ref{fig:oneway} the actual travel time $t_{AB}=t_B-t_A$ from $A$ to $B$, as distinct from the clock indicated travel time $T_{AB}=T_B-T_A$,  is determined by
\begin{equation}
V(v\cos(\theta))t_{AB}=L+v\cos(\theta)t_{AB}
\label{eqn:traveltime1}\end{equation}
where the 2nd term comes from the end $B$ moving an additional distance  $v\cos(\theta)t_{AB}$ during time interval $t_{AB}$.  
With Fresnel drag $V(v)=\frac{c}{n}+v\left(1-\frac{1}{n^2}\right)$, when $V$ and $v$ are parallel, and where $n$ is the dielectric refractive index.   Then
\begin{equation}
t_{AB}= \frac{L}{V(v\cos(\theta))-v\cos(\theta)}=\frac{nL}{c}+\frac{v\cos(\theta)L}{c^2}+..
\label{eqn:traveltime2}\end{equation}
  However if there is no Fresnel drag effect, $V=c/n$, as is the case in RF coaxial cables,  then we obtain 
\begin{equation}
t_{AB}\!=\! \frac{L}{V(v\cos(\theta))-v\cos(\theta)}\!=\!\frac{nL}{c}\!+\frac{v\cos(\theta)Ln^2}{c^2}+..
\label{eqn:traveltime2b}\end{equation}
It would appear that the two terms in (\ref{eqn:traveltime2}) or (\ref{eqn:traveltime2b}) can be separated by rotating the apparatus, giving the magnitude and direction of $\bf v$.
However  it is $T_{AB}=T_B-T_A$ that is measured, and not $t_{AB}$, because of an unknown fixed clock offset $\tau$, as the clocks are not {\it a priori} synchronised,  and as well  an angle dependent clock transport offset $\Delta \tau$, at least until we can establish clock synchronisation, as explained below.  Then the clock readings are $T_A=t_A$ and $T_B=t_B+\tau$, and $T^\prime_B=t^\prime_B+ \tau  +\Delta \tau$, where $\Delta \tau$ is a clock offset that arises from slowing of  clock $C_2$ as it is transported during the rotation through angle $\Delta \theta$, see 
Fig.\ref{fig:oneway}.
\section{Clock Transport Effect}
 The clock transport offset $\Delta \tau$ follows from the clock motion effect
\begin{equation}
\Delta\tau=dt\sqrt{1-\frac{(\bf{v}+\bf{u})^2}{c^2}}-dt\sqrt{1-\frac{\bf{v}^2}{c^2}}= -dt\frac{{\bf v}\cdot{\bf u}}{c^2}+... ,
\label{eqn:offset}\end{equation} 
when clock $C_2$ is transported at velocity $\bf u$ over time interval $dt$, compared to $C_1$.   Now  ${\bf v}\cdot{\bf u}=vu\sin(\theta)$ and $dt= L\Delta \theta/u$. Then the change  in $T_{AB}$ from  this small rotation is,  using (\ref{eqn:traveltime2}) for the case of no Fresnel drag,
\begin{equation}
\Delta T_{AB} =+\frac{v\sin(\theta)Ln^2\Delta \theta}{c^2}-\frac{v\sin(\theta)L\Delta \theta}{c^2}+...
\label{eqn:DeltaT}\end{equation}
as the clock transport effect appears to make the clock-deter-mined travel time shorter (2nd term). Integrating we get 
\begin{equation}
T_B-T_A =\frac{nL}{c}+\frac{v\cos(\theta)L(n^2-1)}{c^2}+\tau,
\label{eqn:TBTA}\end{equation}
where $\tau$ is now the constant offset time. The $v\cos(\theta)$ term may be separated by means of the angle dependence.   Then the value of $\tau$ may be determined, and the clocks synchronised. However if the propagation medium is  vacuum, liquid, or dielectrics such as glass and optical fibres, the Fresnel drag effect is present, and we then use  (\ref{eqn:traveltime1}), and not (\ref{eqn:traveltime2}). Then in   (\ref{eqn:TBTA}) we need make the replacement $n\rightarrow1$, and then the 1st order in $v/c$  term  vanishes, as reported by Spavieri {\it et al.}.  However, in principle, separated clocks may be synchronised by using RF coaxial cables.

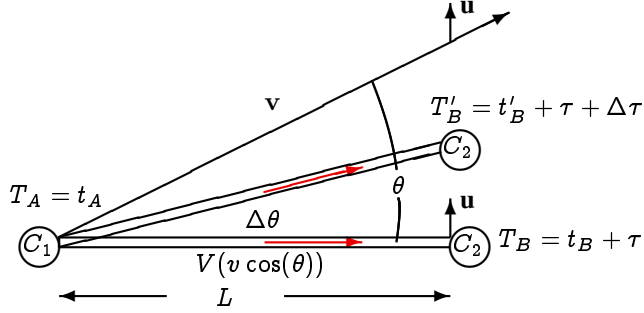
\begin{figure}
\vspace{31mm}
\hspace{-11mm}
\setlength{\unitlength}{1.3mm}
\hspace{0mm}\begin{picture}(0,0)
\thicklines

\definecolor{hellgrau}{gray}{.8}
\definecolor{dunkelblau}{rgb}{0, 0, .9}
\definecolor{roetlich}{rgb}{1, .7, .7}
\definecolor{dunkelmagenta}{rgb}{.9, 0, .0}
\definecolor{green}{rgb}{0, 1,0.4}
\definecolor{black}{rgb}{0, 0, 0}

\color{dunkelmagenta}
\put(30,5.5){\vector(1,0){10}}
\put(30,10.7){\vector(4,1){10}}

\color{black}

\put(9,5){\line(1,0){40}}
\put(9,6){\line(1,0){40}}

\put(9,5){\line(0,1){1}}
\put(49,5){\line(1,1){1}}

\put(4,9.5){{$T_A=t_A$}}

\put(54,5){{$T_B=t_B+\tau$}}
\put(47,18.5){{$T^\prime_B=t^\prime_B+\tau+\Delta \tau$}}
\put(50,15){\circle{4}}
\put(48.3,14.5){{$C_2$}}

\put(49,6){\vector(0,1){4}}
\put(50,9){\bf{u}}

\put(9,6){\vector(2,1){46}}
\put(9,6){\line(4,1){39}}
\put(9,5){\line(4,1){39}}
\put(48,14.5){\line(1,1){1}}
\put(49,26){\vector(0,1){4}}
\put(50,29){\bf{u}}

\put(30,19){{${\bf v}$}}
\put(28,7){{$\Delta\theta$}}

\put(43,10.5){{$\theta$}}

\qbezier(43.5,13)(43,17)(41,22)
\qbezier(43.5,5.6)(44,9)(43.7,10)

\put(23,2.5){{$V(v\cos(\theta))$}}

\put(7,5){\circle{4}}
\put(5.3,4.5){{$C_1$}}

\put(51,5){\circle{4}}

\put(49.3,4.5){{$C_2$}}

\put(20,0){\vector(-1,0){11}}
\put(25,-1){$ L$}
\put(32,0){\vector(1,0){17}}

\end{picture}

\vspace{0mm}
	\caption{\small{ Schematic layout for measuring the one-way speed of light in either free-space, optical fibres or RF coaxial cables, without requiring the synchronisation of the clocks $C_1$ and $C_2$. Here $\tau$ is the, initially  unknown, offset time between the clocks. Times $t_A$ and $t_B$ are true times, without clock offset and clock transport effects, while $T_A=t_A$, $T_B=t_A+\tau$ and $T^\prime_B=t^\prime_B+\tau+\Delta \tau$ are clock readings. $V(v\cos(\theta))$ is the speed of EM radiation wrt the apparatus before rotation, and   $V(v\cos(\theta-\Delta\theta))$ after rotation,  ${\bf v}$ is the velocity of the apparatus through space in direction $\theta$ relative to the apparatus before rotation, $\bf u$ is the velocity of transport for clock $C_2$,  and $\Delta \tau<0$ is the net slowing of clock $C_2$ from clock transport, when apparatus is rotated through angle $\Delta \theta>0$.  Note that ${\bf v}\cdot{\bf u}>0$.}}
 \label{fig:oneway}
\end{figure}

\section{DeWitte 1st Order in $\bf v/c$ Detector}
 The DeWitte  $L=1.5$km 5MHz RF coaxial cable experiment, Brussels 1991,   was a double 1st order in $v/c$ detector, using the scheme in Fig.\ref{fig:oneway},  but employing a 2nd RF coaxial cable   for the opposite direction, giving clock difference $T_D-T_C$, to cancel temperature effects, and also used   3 Caesium  atomic clocks at each end.   The orientation was NS and rotation was achieved by that of the earth \cite{DeWitte}.    Then
 \begin{equation}
 T_{AB}-T_{CD}=\frac{2v\cos(\theta)L(n^2-1)}{c^2}+2\tau
 \label{eqn:DeWitte}\end{equation}
 For a horizontal detector the dynamic range of $\cos(\theta)$ is \newline $2\sin(\lambda)\cos(\delta)$, caused by the earth rotation, where $\lambda$ is the latitude of the detector location and $\delta$ is the declination of $\bf v$.  The value of $\tau$ may be determined   and the clocks synchronised.  Some of DeWiite's data and results are in Figs. \ref{fig:Coax2009} and \ref{fig:DewItteSidereal}.
 \begin{figure}[t]
 \vspace{-15mm}
 \hspace{3.5mm}\includegraphics[scale=0.525]{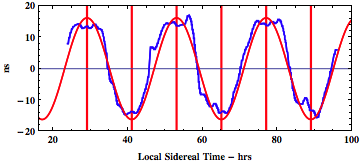}
 
 \hspace{-3mm}\includegraphics[scale=0.22]{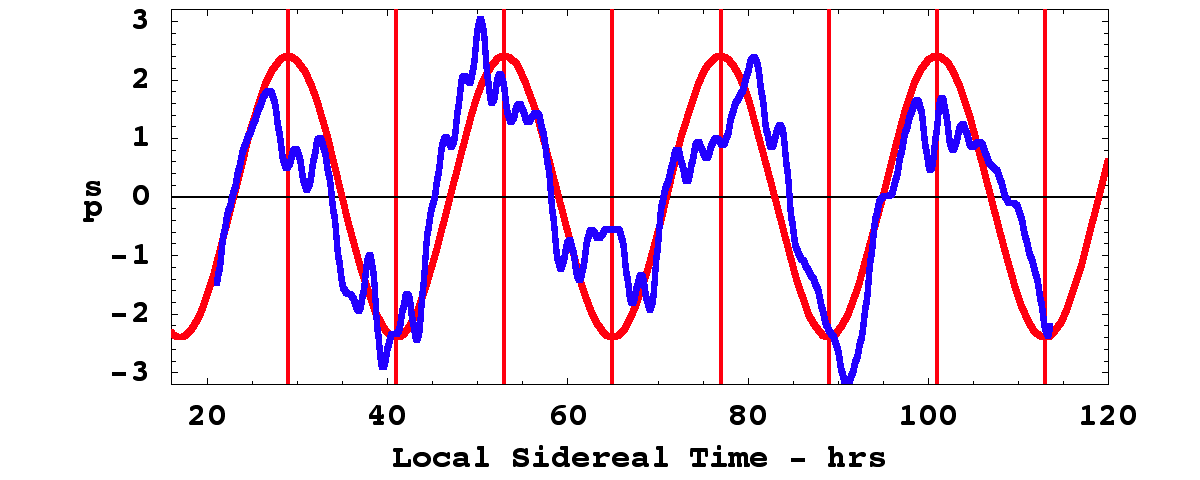}
\vspace{-7mm}	\caption{\small {Top:  Data from the 1991 DeWitte NS RF coaxial cable experiment, $L=1.5$km, using the arrangement shown in Fig.\ref{fig:oneway}, with a 2nd RF coaxial cable carrying a signal in the reverse direction.  The vertical red lines are at RA=5$^{h}$. DeWitte gathered data for 178 days, and showed that the crossing time tracked sidereal time, and not local solar time, see Fig.\ref{fig:DewItteSidereal}.  DeWitte reported that $v \approx 500km/s$. If a Fresnel drag effect is included  no effect would have been seen.  Bottom:   Dual coaxial cable detector  data  from May  2009 using the technique in Fig.\ref{fig:DualCoax} with $L=20$m. NASA Spacecraft  Earth-flyby Doppler shift data predicts Dec $=-77^\circ$, $v=480$km/s, giving a sidereal dynamic range of 5.06ps, very close to that observed.  The  vertical red lines are at RA=5$^{h}$. In both data sets we see the earth sidereal rotation effect together with significant wave/turbulence effects.}}
\label{fig:Coax2009}\end{figure}
 \begin{figure}[h]
 \vspace{-3mm}
 \hspace{3.5mm}\includegraphics[scale=0.5]{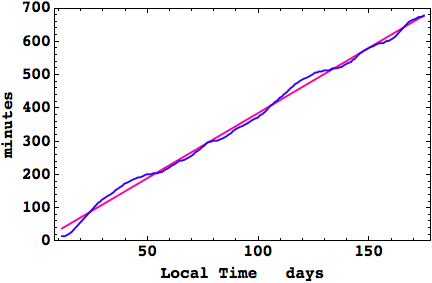} 
\vspace{-2mm}	\caption{\small {DeWitte  collected data over 178 days and demonstrated that the zero crossing time, see Fig.\ref{fig:Coax2009}, tracked sidereal time and not local solar time. The plot shows the negative  of the drift  in the crossing time vs local solar time, and has a slope, determined by the best-fit straight line, of -3.918 minutes per day, compared to the actual average value of  -3.932 minutes per day. Again we see fluctuations from day to day.} }
\label{fig:DewItteSidereal}\end{figure}

  \begin{figure}
\vspace{5mm}
\setlength{\unitlength}{1.1mm}

\vspace{20mm}

\hspace{4mm}\begin{picture}(0,0)
\thicklines

\definecolor{hellgrau}{gray}{.8}
\definecolor{dunkelblau}{rgb}{0, 0, .9}
\definecolor{roetlich}{rgb}{1, .7, .7}
\definecolor{dunkelmagenta}{rgb}{.9, 0, .0}
\definecolor{mauve}{rgb}{0.4, 0, .8}

\put(0,6.5){\bf S}\put(65,6.5){\bf N}

\put(5.5,-2){\bf A}
\put(67,-2){\bf B}

\put(5.5,16){\bf D}
\put(67,16){\bf C}

\put(0,0){\line(0,1){5}}\put(0,0){\line(1,0){5}}
\put(0,5){\line(1,0){5}}\put(5,0){\line(0,1){5}}
\put(0.2,1.9){\bf RF}

\put(0,10){\line(0,1){5}}\put(0,10){\line(1,0){5}}

\put(0,15){\line(1,0){5}}\put(5,10){\line(0,1){5}}
\put(0.4,12.6){\bf DS}
\put(0.6,10.2){\bf O}

\put(20,-5){\vector(-1,0){15}}
\put(25,-5.5){\large$ L$}
\put(32,-5){\vector(1,0){37}}

 \color{dunkelmagenta}
  \put(32,-2){\bf FSJ1-50A}
  \put(32,8.0){\bf FSJ1-50A}
\put(5,1.1){\line(1,0){64}}
\put(35,1.21){\vector(1,0){1}}
\put(35,11.3){\vector(-1,0){1}}
 
\color{dunkelblau}
  \put(32,15){\bf HJ4-50}
    \put(32,5){\bf HJ4-50}
\put(5,4){\line(1,0){64}}
\put(5,3.6){\line(1,0){64}}

  \color{dunkelmagenta}
\put(5,11.4){\line(1,0){64}}

 \color{dunkelblau}
\put(5,14){\line(1,0){64}}
\put(5,13.6){\line(1,0){64}}

\put(35,13.8){\vector(-1,0){1}}

\put(35,3.9){\vector(1,0){1}}

 \color{dunkelmagenta}
\put(68.4,8.92){\oval(5,10.1)[rb]}
\put(68.4,6.3){\oval(5,10.1)[rt]}

\put(68.5,6.2){\oval(10,10.1)[rb]}
\put(68.5,8.8){\oval(10,10)[rt]}
\put(73.5,6){\line(0,1){4}}
\put(73.6,6){\vector(0,1){3}}
\put(70.9,6.0){\vector(0,1){2}}

\end{picture}

\vspace{5mm}\caption{\small Because Fresnel drag is absent in RF coaxial cables this dual cable setup, using one clock (10MHz RF source and   Digital Storage Oscilloscope DSO to measure and store timing difference between the two circuits, as in (\ref{eqn:RF2})), is capable of detecting the absolute motion of the detector  wrt to space, revealing the earth sidereal rotation effect as well as wave/turbulence effects. Results from such an experiment are shown in Fig.\ref{fig:Coax2009}. Andrews phase-stablised coaxial cables are used. More recent results are reported in \cite{CahillGW2}. }
\label{fig:DualCoax}
\end{figure}
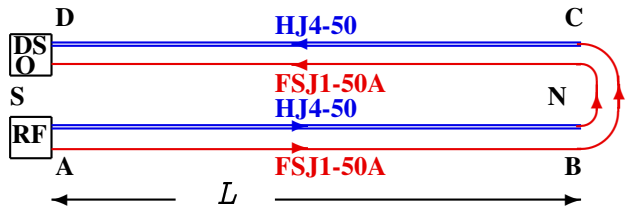

\section{Dual RF Coaxial Cable Detector}
The single clock Dual RF Coaxial Cable  Detector exploits the absence of the Fresnel drag effect in RF coaxial cables \cite{CahillGW2}.  Then from  (\ref{eqn:traveltime2b})   the round trip  travel time for one circuit is, see  Fig.\ref{fig:DualCoax},
\vspace{0mm}
\begin{equation}
t_{AB}+t_{CD}=\frac{(n_1+n_2)L}{c}+\frac{v\cos(\theta)L(n_1^2-n_2^2)}{c^2}+..
\label{eqn:RF1}\end{equation}
where $n_1$ and $n_2$ are the  effective refractive indices for the two  different RF coaxial cables.  There is no clock transport effect as the detector is rotated. Dual circuits  reduce temperature effects.    The travel time difference of the two circuits at the DSO is then
 \begin{equation}
\Delta t=\frac{2v\cos(\theta)L(n_1^2-n_2^2)}{c^2}+..
\label{eqn:RF2}\end{equation}
A sample of data is shown in Fig.\ref{fig:Coax2009}, and is in excellent agreement with the DeWitte data, the NASA flyby Doppler shift data, and the Michelson-Morley and Miller results.

\section{Conclusions}
The absence of the Fresnel drag in RF coaxial cables enables  1st order in $v/c$ measurements of the anisotropy of the speed of light.  DeWitte pioneered  this using the multiple clock technique, and took account of the clock transport effect, while the new dual RF coaxial cable detector uses only one clock.  This provides a very simple and robust technique.   Experiments by Michelson and Morley 1887, Miller 1925/26,  DeWitte 1991, Cahill 2006, 2009, 2012, and NASA earth-flyby Doppler shift data now all agree, giving the solar system a speed of $\sim486$km/s in the direction RA=4.3$^h$, Dec=-75.0$^{\circ}$.   These experiments have detected the fractal textured  dynamical structure of space - the privileged local frame \cite{CahillGW2}. This report is from the Gravitational Wave Detector Project at Flinders University.

\small{

}

\end{document}